\documentclass[letterpaper,11pt]{article}
\usepackage{jheppub}
\usepackage{graphicx}
\usepackage{braket}
\usepackage{tikz}
\usepackage{hyperref}
\usepackage{amssymb}
\usepackage{amsmath}
\usepackage{color}
\usepackage{zx-calculus}
\usepackage{adjustbox}
\usepackage{tikzit}
\input{zx.tikzdefs}

\tikzstyle{gn}=[draw=black, shape=circle, fill={zx_green}, draw=black, inner sep=0.7mm, outer sep=-0.2em, minimum width=0pt, minimum height=0pt, tikzit fill={rgb,255: red,181; green,215; blue,181}]
\tikzstyle{rn}=[gn, fill={zx_red}, draw=black, tikzit fill={rgb,255: red,215; green,96; blue,96}]
\tikzstyle{fn}=[gn, fill={rgb,255: red,255; green,220; blue,185}, tikzit fill={rgb,255: red,255; green,220; blue,185}, tikzit draw=black]
\tikzstyle{gn_phase}=[shape=rectangle, fill={zx_green}, draw=black, minimum size=1.2em, rounded corners=0.5em, inner sep=0.25em, outer sep=-0.2em, scale=0.8, font={\footnotesize}, tikzit shape=circle, tikzit fill={rgb,255: red,181; green,215; blue,181}]
\tikzstyle{rn_phase}=[{gn_phase}, fill={zx_red}, draw=black, tikzit fill={rgb,255: red,215; green,96; blue,96}]
\tikzstyle{fn_phase}=[{gn_phase}, fill={rgb,255: red,255; green,220; blue,185}, tikzit fill={rgb,255: red,240; green,207; blue,174}, tikzit draw=black]
\tikzstyle{hbox}=[box, draw=black, shape=rectangle, fill=yellow, minimum size=.67em, font={\tiny}]
\tikzstyle{hdot}=[fill=yellow, draw=black, shape=rectangle, inner sep=0.6mm, minimum height=1.5mm, minimum width=1.5mm]
\tikzstyle{z_sheet_label}=[scale=0.8, font={\footnotesize}, tikzit shape=circle, tikzit fill={rgb,255: red,181; green,200; blue,177}, tikzit draw={rgb,255: red,180; green,180; blue,180}, fill={rgb,255: red,200; green,200; blue,200}, shape=rectangle, inner sep=2pt, rounded corners=0.3em, fill opacity=0.4, text opacity=1]
\tikzstyle{x_sheet_label}=[scale=0.8, font={\footnotesize}, tikzit shape=circle, tikzit fill={rgb,255: red,200; green,169; blue,169}, tikzit draw={rgb,255: red,180; green,180; blue,180}]
\tikzstyle{box}=[draw=black, shape=rectangle, fill=white, minimum size=.70em, inner sep=0.10em, scale=0.85, font={\tiny}]
\tikzstyle{small box}=[fill=white, draw=black, shape=rectangle, minimum width=0.75cm, minimum height=0.75cm]
\tikzstyle{long box}=[fill=white, draw=black, shape=rectangle, minimum width=0.75cm, minimum height=1.25cm]
\tikzstyle{very long box}=[fill=white, draw=black, shape=rectangle, minimum width=0.75cm, minimum height=1.75cm]
\tikzstyle{white square}=[fill=white, draw=black, shape=rectangle, font={\LARGE}]
\tikzstyle{left_text}=[anchor=west, tikzit shape=circle]
\tikzstyle{graphv}=[fill=white, draw=none, shape=circle, inner sep=0mm]
\tikzstyle{medpbox}=[shape=squareNE, draw=black, fill=lightgray, inner sep=0.5mm, minimum height=12mm, minimum width=4mm, font={\footnotesize}, tikzit shape=rectangle]
\tikzstyle{longpbox}=[shape=squareNE, draw=black, fill=lightgray, inner sep=0.5mm, minimum height=18mm, minimum width=4mm, font={\footnotesize}, tikzit shape=rectangle]
\tikzstyle{dash_circ}=[shape=circle, draw=black, dashed, inner sep=0.7mm, minimum width=3mm, minimum height=3mm, tikzit fill={rgb,255: red,128; green,128; blue,128}]
\tikzstyle{black dot}=[fill=black, draw=black, shape=circle, inner sep=1pt]
\tikzstyle{label}=[font={\scriptsize}]

\tikzstyle{arrow}=[->]
\tikzstyle{mid_arrow}=[->]
\tikzstyle{black_brace}=[-, decorate, decoration={brace, amplitude=2mm, raise=-1mm}]
\tikzstyle{grey_brace}=[-, draw=gray, decorate, decoration={brace, amplitude=2mm, raise=-1mm}]
\tikzstyle{blue_edge}=[-, very thick, draw=blue]
\tikzstyle{hadamard_edge}=[-, dashed, draw=blue]
\tikzstyle{dashed}=[-, dash pattern=on 0.6mm off 0.5mm]
\tikzstyle{grey_dashed}=[-, draw=gray, dash pattern=on 0.8mm off 0.8mm]
\tikzstyle{blue}=[-, draw=blue]
\tikzstyle{thick}=[-, line width=1pt]
\tikzstyle{dotsedge}=[-, dotted, decoration={brace, amplitude=2mm, raise=-1mm}]
\tikzstyle{classical}=[-, double, double distance=0.75pt]

\DeclareMathOperator{\tr}{tr}

\newcommand{\nn}{\\ \nonumber}

\newcommand \mathtikz[1] {\quad \vcenter{\hbox{\tikz{#1}}} \quad}

\usepackage{tikz}
\usetikzlibrary{decorations.markings, arrows.meta}

\newcommand\stripRR[2]{ 
    \begin{scope}[xshift=#1,yshift=#2]
        \draw[
            line width=1.5pt,
            decoration={
                markings,
                mark=at position 0.55 with {\arrow[line width=1.5pt, scale=1]{Stealth}}
            },
            postaction={decorate}
        ] (-0.5,0) -- (-0.5,-2);
        \draw[
            line width=1.5pt,
            decoration={
                markings,
                mark=at position 0.55 with {\arrow[line width=1.5pt, scale=1]{Stealth}}
            },
            postaction={decorate}
        ] (0.5,-2) -- (0.5,0);
        \draw[dashed, line width=1pt] (-0.5,-1) -- (0.5,-1);
        \node[above, font=\small] at (-0.5,0) {$a$};
        \node[above, font=\small] at (0.5,0) {$\bar a$};
    \end{scope}
}

\newcommand{\openstrings}[4][2.5]{%
    \begin{tikzpicture}[baseline=-0.5ex]
        \def\sep{0.55}
        \def\wid{#1}
        \tikzset{
            string/.style={
                thick,
                decoration={
                    markings,
                    mark=at position 0.55 with {\arrow{stealth}}
                },
                postaction={decorate}
            }
        }
        \foreach \k in {1,...,#2}{
            \pgfmathsetmacro{\y}{-(\k-1)*\sep}
            \draw[string] (0,\y) -- (\wid,\y);
            \node[left, font=\small] at (0,\y) 
                {$#3_{\k}$};
            \node[right, font=\small] at (\wid,\y) 
                {$#4_{\k}$};
        }
        \pgfmathsetmacro{\ydots}{-(#2)*\sep}
        \node at (\wid/2, \ydots) {$\vdots$};
        \pgfmathsetmacro{\ybot}{-(#2+1)*\sep}
        \draw[string] (0,\ybot) -- (\wid,\ybot);
        \node[left, font=\small] at (0,\ybot) 
            {$#3_{n}$};
        \node[right, font=\small] at (\wid,\ybot) 
            {$#4_{n}$};
    \end{tikzpicture}%
}

\newcommand\idA[2]{ 
    \begin{scope}[xshift=#1,yshift=#2]
        \filldraw[fill=white,draw=black] (-0.25,0) rectangle (0.25,-1);
        \draw[
            decoration={
                markings,
                mark=at position 0.55 with {\arrow{stealth}}
            },
            postaction={decorate}
        ] (-0.25,0) -- (0.25,0);
        \draw[
            decoration={
                markings,
                mark=at position 0.55 with {\arrow{stealth}}
            },
            postaction={decorate}
        ] (-0.25,-1) -- (0.25,-1);
    \end{scope}
}

\newcommand\muC[2]{ 
\begin{scope}[xshift=#1,yshift=#2]
\filldraw[left color=lightgray, right color=white] (-0.25,0) to [out=-90,in=180] (0,-0.33) to [in=-90,out=0] (0.25,0) to  (0.75,0) to [in=90,out=-90] (0.25,-1) to [out=-90,in=-90] (-0.25,-1) to [in=-90,out=90] (-0.75,0);
\filldraw[left color=white,right color=lightgray] (-0.5,0) ellipse (0.25 and 0.1);
\filldraw[left color=white,right color=lightgray] (0.5,0) ellipse (0.25 and 0.1);
\draw[dotted] (0.25,-1) arc (0:180:0.25 and 0.1);
\end{scope}
}

\newcommand\deltaC[2]{
\begin{scope}[xshift=#1,yshift=#2]
\filldraw[left color=lightgray, right color=white] (-0.25,-1) to [out=90,in=180] (0,-0.66) to [in=90,out=0] (0.25,-1) to [out=-90,in=-90] (0.75,-1) to [in=-90,out=90] (0.25,0) to (-0.25,0) to [in=90,out=-90] (-0.75,-1) to [out=-90,in=-90] (-0.25,-1);
\filldraw[left color=white,right color=lightgray] (0,0) ellipse (0.25 and 0.1);
\draw[dotted] (-0.25,-1) arc (0:180:0.25 and 0.1);
\draw[dotted] (0.75,-1) arc (0:180:0.25 and 0.1);
\end{scope}
}

\newcommand\muA[2]{ 
\begin{scope}[xshift=#1,yshift=#2]
\draw (-0.75,0) -- (-0.25,0) to [out=-90,in=180] (0,-0.33) to [in=-90,out=0] (0.25,0) -- (0.75,0) to [in=90,out=-90] (0.25,-1);
\draw (-0.25,-1) -- (0.25,-1);
\draw (-0.75,0) to [in=90,out=-90] (-0.25,-1);
\end{scope}
}

\newcommand\nablaA[2]{ 
\begin{scope}[xshift=#1,yshift=#2]
\draw (-0.25,-0.5) -- (0.25,0) -- (0.75,0) -- (0.25,-0.5) --(0.25,-1) -- (-0.25,-1) -- (-0.25,-0.5) -- (-0.75,0) -- (-0.25,0) -- (0,-0.25);
\draw[dashed] (-0.25,-0.5) -- (0.25,-0.5);
\end{scope}
}

\newcommand\deltaA[2]{ 
\begin{scope}[xshift=#1,yshift=#2]
\draw (-0.75,-1) -- (-0.25,-1) to [out=90,in=180] (0,-0.66) to [in=90,out=0] (0.25,-1) -- (0.75,-1) to [in=-90,out=90] (0.25,0) -- (-0.25,0) to [in=90,out=-90] (-0.75,-1);
\end{scope}
}

\newcommand\zipper[2]{ 
\begin{scope}[xshift=#1,yshift=#2]
\draw (-0.25,-1) -- (0.25,-1);
\filldraw[right color=white,left color=lightgray] (-0.25,0) to (-0.25,-1) to [out=90,in=225] (0,-0.5) to [out=-45,in=90] (0.25,-1) to (0.25,0);
\filldraw[left color=white,right color=lightgray] (0,0) ellipse (0.25 and 0.1);
\end{scope}
}

\newcommand\cozipper[2]{ 
\begin{scope}[xshift=#1,yshift=#2]
\draw (-0.25,0) -- (0.25,-0);
\filldraw[right color=white,left color=lightgray] (-0.25,-1) to (-0.25,0) to [out=-90,in=135] (0,-0.5) to [out=45,in=-90] (0.25,0) to (0.25,-1) to [in=-90,out=-90] (-0.25,-1);
\draw[dotted] (0.25,-1) arc (0:180:0.25 and 0.1);
\end{scope}
}

\newcommand\epsilonC[2]{ 
\begin{scope}[xshift=#1,yshift=#2]
\filldraw[right color=white,left color=lightgray] (-0.25,0) to [out=-90,in=180] (0,-0.33) to [in=-90,out=0] (0.25,0);
\filldraw[left color=white,right color=lightgray] (0,0) ellipse (0.25 and 0.1);
\end{scope}
}

\newcommand\etaC[2] { 
\begin{scope}[xshift=#1,yshift=#2]
\filldraw[right color=white,left color=lightgray] (-0.25,0) to [out=90,in=180] (0,0.33) to [in=90,out=0] (0.25,0) to [in=-90,out=-90] (-0.25,0);
\draw[dotted] (0.25,0) arc (0:180:0.25 and 0.1);
\end{scope}
}

\newcommand\epsilonA[2] {
\begin{scope}[xshift=#1,yshift=#2]
\draw (-0.25,0) -- (0.25,0);
\draw (-0.25,0) to [out=-90,in=180] (0,-0.33) to [in=-90,out=0] (0.25,0);
\end{scope}
}

\newcommand\FrobetaA[2] {
\begin{scope}[xshift=#1,yshift=#2]
\draw (-0.25,0) -- (0.25,0);
\draw (-0.25,0) to [out=90,in=180] (0,0.33) to [in=90,out=0] (0.25,0);
\end{scope}
}

\begin{document}
\author[a]{Gabriel Wong}
\emailAdd{gabrielwon@gmail.com}
\affiliation[a]{Mathematical Institute, University of Oxford, Andrew Wiles Building, Radcliffe Observatory Quarter, Woodstock Road, Oxford, OX2 6GG, U.K.}
\author[b]{Razin A. Shaikh}
\affiliation[b]{Department of Computer Science, University of Oxford, Wolfson Building, Parks Road, Oxford, OX1 3QD, U.K.}
\title{A ribbon ZX calculus for gauge theory}
\author[]{William Donnelly}
\emailAdd{williamdonnelly@gmail.com}
\abstract{ZX-calculus provides a graphical formalism for reasoning about quantum processes, built from two interacting Frobenius algebras associated with the Z and X bases of a qubit.    While it has found widespread application in quantum information and computing, its relationship to quantum field theory has only recently begun to be explored. In this work, we further develop this connection by providing a generalization of ZX-calculus to two-dimensional Yang–Mills theory with a compact gauge group.  The key observation is that both frameworks can be organized around the Hopf-Frobenius algebraic structure associated with a group algebra, which can in turn be described by the diagrammatics of two dimensional  topological quantum field theory.   Given the well known relationship between gauge theory and gravity in two and three dimensions,  our work paves the way for applications of ZX to low dimensional gravity.
}

\maketitle
\section{Introduction}
Quantum mechanics is traditionally formulated in terms of the linear algebra of complex vector spaces, endowed with vector addition and scalar multiplication. However, an efficient description of quantum information processes involves operations like entangling and copying, which correspond to multiplication and co-multiplications of quantum states.
ZX-calculus~\cite{coeckeInteractingQuantumObservables2008, coeckeInteractingQuantumObservables2011} provides a graphical formalism that incorporates these operations into a diagrammatical calculus for reasoning about qubit quantum processes.  It is built from the two ``spiders" describing two product structures on the Hilbert space of a qubit. 
\begin{align}
    \input{x-spider-mult.tikz} \qquad \qquad  \input{z-spider-mult.tikz}
\end{align}
They define two strongly complementary Frobenius algebras, conventionally associated with the 
Z and X bases of a qubit. The key advantage of ZX-calculus over a generic tensor network is a concise set of rewrite rules for simplifying complex algebraic expressions. Enriching the spiders with phases leads to a universal and complete language for qubit quantum mechanics, so that any valid relation between qubit linear maps can be derived from ZX rewrites ~\cite{ngUniversalCompletionZXcalculus2017, vilmartNearMinimalAxiomatisationZXCalculus2019}.
Since the original work on qubits, ZX-calculus and its completeness have been extended to finite~\cite{poorCompletenessArbitraryFinite2023, poorZXcalculusComplete2024, wangCompletenessQufiniteZXW2024}, and infinite-dimensional settings~\cite{shaikhFockedupZX2024}.

ZX-calculus has become a widely used tool in quantum information and quantum computing, with applications ranging from quantum circuit optimization~\cite{debeaudrapFastEffective2020,duncanGraphtheoreticSimplificationQuantum2020, kissingerReducingTcountZXcalculus2020}, quantum error correction~\cite{tanSATScalpel2024, huangGraphicalCSSCode2023, teagueFloquetifyingColourCode2023}, fault-tolerance~\cite{rodatzFaultTolerance2025, poor2025ultralowoverheadsyndrome, bauer2025planar}, measurement-based quantum computation~\cite{kissingerUniversalMBQCGeneralised2019, BackensPerezPlanarMeasured2025, kissinger2026zxflow}, quantum chemistry~\cite{shaikhHowSum2023, defeliceLightMatterInteractionZXW2023, mcdowallrose2025fromfermions}, quantum machine learning~\cite{wangDifferentiatingIntegrating2024}, and condensed matter physics~\cite{eastAKLTStatesZXDiagramsDiagrammatic2022, lu2026generalized, mendoza2025graphical}.
More recently, it has made contact with high energy physics: it elucidates the categorical structure of Feynman diagrams~\cite{shaikhCategoricalSemanticsFeynman2022} and can be used to perform calculations on spin networks in loop quantum gravity~\cite{wang2025beyond, eastSpinnetworksZXcalculus2022}.  These connections point toward a deeper relationship between ZX-calculus and the algebraic structures of quantum field theory.

In this work, we further develop this connection by providing a generalization of ZX-calculus to two-dimensional gauge theory, and in particular to two-dimensional Yang--Mills with a compact gauge group.\footnote{Our results also hold for more general two-dimensional topological gauge theories, such as BF theory.} The key observation is that both 2D Yang--Mills and ZX-calculus admit a description in terms of Frobenius algebras --- the same algebraic language that underlies the categorical approach to two-dimensional topological field theory.  This connection makes ZX rewrite rules manifest as topological equivalences, placing the calculus within the broader framework of 2D TQFT. More precisely, we employ the framework of area-dependent QFT~\cite{Runkel:2018uls}, which accommodates the infinite-dimensional Hilbert spaces that arise in 2D Yang--Mills while retaining the familiar features of TQFT. As we discuss in the conclusion, this work provides a starting point for exploring a broader relationship between ZX-calculus and low-dimensional gauge theory and gravity.

The influence of TQFT on ZX-calculus goes back to its origins. Frobenius algebras  play a central role in categorical quantum mechanics (CQM)~\cite{AbramskyCoecke2004}, which formulates quantum theory in the language of symmetric monoidal categories. 
This categorical viewpoint comes equipped with the graphical language of tensor networks~\cite{JoyalStreet1991}, allowing quantum processes and their equations to be represented diagrammatically. 
In finite-dimensional Hilbert spaces, special commutative dagger Frobenius algebras give a categorical description of orthonormal bases~\cite{CoeckePavlovicVicary2012}.
The study of complementary Frobenius algebras in this setting led to the development of the ZX-calculus~\cite{coeckeInteractingQuantumObservables2008, coeckeInteractingQuantumObservables2011}.

Despite their common algebraic foundation in Frobenius algebras, the diagrammatic languages used in two-dimensional TQFT and in the ZX-calculus look quite different.
The former is expressed in terms of 2D cobordisms, while the latter is formulated in terms of tensor diagrams with 1D wires.
In this paper, we make the relationship between these two diagrammatic traditions explicit. 
The key observation is that both frameworks can be organized around the Hopf-Frobenius algebraic structure associated with a group algebra. 
Using this perspective, we develop a parallel dictionary between ZX diagrams and TQFT diagrams. 
In particular, we introduce decorations of TQFT diagrams that play the role of ZX phases, thereby importing the phase group structure of the ZX-calculus into the two-dimensional TQFT setting. 
This construction provides a natural continuum generalization of the ZX-calculus in two dimensional gauge theories.

 We will begin by giving a brief overview of 2DYM as an area dependent TQFT, and formulate a generalization of the usual TQFT diagrammatics.  The diagrams take the form of ribbon graphs with trivial braiding.  We give two useful interpretations of these diagrams, first as worldsheets for stacks of open strings, and second as worldlines of entangled anyons.  We then provide a dictionary relating these ribbon graphs to conventional ZX diagrams. We show how the basic relations in ZX become topologically manifest when interpreted in terms of string worldsheets and entangled anyons.   In the conclusion,  we will comment on q-deformations, and the large N limit.

\section{Two-dimensional Yang-Mills as an area-dependent TQFT}
Two-dimensional Yang-Mills theory is a non-abelian gauge theory. The fundamental field is a gauge potential $A_\mu$, a Lie-algebra valued one-form on a two-dimensional spacetime manifold $M$. Its curvature is the field strength
\begin{equation}
F_{\mu\nu} = \partial_\mu A_\nu - \partial_\nu A_\mu + [A_\mu, A_\nu],
\end{equation}
which measures the failure of parallel transport around infinitesimal loops. The dynamics are governed by the action
\begin{equation}
I = \frac{1}{4 g_\text{YM}^2} \int_M \sqrt{g}\, \tr[ F^{\mu \nu} F_{\mu \nu}].
\end{equation}
The theory is invariant under gauge transformations $A_\mu \to A_\mu + \partial_\mu f$. A key additional feature is invariance under area-preserving diffeomorphisms: this becomes manifest when we write the action in terms of differential forms. Then only the volume form appears rather than the full metric, so the theory depends only on the total area of $M$ and its topology. This makes 2DYM an \emph{area-dependent} topological field theory \cite{Runkel:2018uls}. It is a generalization of a TQFT, where the the Hilbert space is infinite dimensional. 

The gauge-invariant content of the theory is captured by the holonomy
\begin{equation}\label{U}
U = \mathcal{P}\exp\left(i\int_0^L dx\, A_x\right) \in G,
\end{equation}
the path-ordered exponential of the gauge field along a spatial slice of length $L$, which may be either an interval or a circle.  Quantization of the gauge-invariant degrees of freedom gives a wavefunctional $\psi(U)$ depending only on the holonomy.  On the circle, the gauge constraints require these wavefunctions to be class functions  satisfying $\psi(U)= \psi (gUg^{-1})$.   On the interval with a particular choice of boundary conditions, the Hilbert space is $L^2(G)$, the space of  square-integrable functions on $G$\footnote{The fact that $U$ arises as a path-ordered exponential is relevant for defining a local tensor product structure in 2DYM, but will not be important for elucidating the algebraic structure of the gauge theory.} 
.  The inner product is defined by the Haar integral over the group:
\begin{equation}
\langle f | h \rangle = \int dU\, f(U) h(U^{-1}),
\end{equation}

\subsection*{The string basis}

A natural basis for $L^2(G)$ is provided by products of matrix elements of the holonomy \eqref{U} in the fundamental representation $N \times N$ representation. We think of each matrix element $U_{ij}$ as describing a single open string with indices $i,j=1,\cdots N$ labeling its endpoints. A general basis element is then a product
\begin{equation}
U_{IJ} = U_{i_1 j_1} \cdots U_{i_n j_n}, 
\end{equation}
with $n$ ranging from $0$ ( corresponding to the constant function $f(U)=1$) to $\infty$. This describes 
a stack of $n$ open strings. 
\begin{align}
    \openstrings{2}{i}{j}
\end{align}
We represent $L^{2}(G)$ as an interval which is made of a direct sum of such open strings states, over all choices of $n$. The propagator  $e^{-TH}$ on this interval is described by a ribbon: 
\begin{align}
    \mathtikz{\idA{0cm}{0cm}}
\end{align}
we interpret it as a superposition of stacks of open string worldsheets.

According to the Peter-Weyl theorem, $L^{2}(G)$ also admits the decomposition into a direct sum over irreps of the group:
\begin{equation}\label{Peter} 
L^2(G) = \bigoplus_a V_a \otimes V_{\bar{a}}
\end{equation}
$V_{a}$ denotes an irrep of $G$, and $V_{\bar{a}}$ its dual. This decomposition is  a statement about the completeness and orthogonality of the representation basis elements.
\begin{align}
    a_{ij}(U),\quad  i,j =1,\cdots d_{a}
\end{align}
which satisfy
\begin{align}
\int dU \,\, a_{ij}(U) b_{kl}(U^{-1}) =\frac{\delta_{ab} \delta_{il} \delta_{kj}}{d_{a}} 
\end{align}
Concretely, Schur-Weyl duality tells us that the representation matrix elements are obtained by symmetrizing /antisymmetrizing the string basis according to the Young Tableaux for $a$. 

The representation basis is useful because $V_a$ has the interpretation of an \textbf{anyon} --- an object in the representation category of $G$. Thus, \eqref{Peter} implies that $L^2(G)$ is an entangled sum of anyons.  We can thus represent a strip as: 
\begin{align}
\mathtikz{\idA{0}{0}}=\bigoplus_{a} \mathtikz{\stripRR{0cm}{0cm}}
\end{align} 
It is this structure that relates the ribbon graphs of 2DYM to anyon diagrams in category theory. This relation continues to hold when we pass to the $q$-deformed theory, where braiding becomes nontrivial. 

\subsection{2DYM as an Open-closed TQFT}

The partition function of 2DYM on a general 2D manifold with boundaries can be computed exactly.  This is because it satisfies the cutting and gluing rules of an area dependent TQFT. 
The area dependence enters through the propagator, which on a strip of area $A$ carries 
a factor $e^{-g^2 A C_2(R)}$ in each irrep $R$, where $C_2(R)$ is the quadratic Casimir. 
Closed manifolds are generated by gluing the following cobordisms (read from top to bottom) :
\begin{align}
    \mathtikz{\muC{0}{0}} \quad
    \mathtikz{\deltaC{0}{0}} \quad
    \mathtikz{\etaC{0}{0}} \quad
    \mathtikz{\epsilonC{0}{0}} 
\end{align}
These are linear maps between the Hilbert spaces of incoming and outgoing circles, with the pair of pants defining a product.   This forms the closed Frobenius algebra. 
Open manifolds have codimension-1 boundaries given by intervals. These map most directly 
into conventional ZX calculus and are generated by the open cobordisms: 
  \begin{align}
    \mathtikz{\muA{0}{0}} \quad
    \mathtikz{\deltaA{0}{0}} \quad
    \mathtikz{\FrobetaA{0}{0}} \quad
    \mathtikz{\epsilonA{0}{0}} 
\end{align}
These form the open Frobenius algebra $\mathcal{A}$. The left-most diagram is the convolution product on
$L^2(G)$ and describes the fusion of open strings along one of their endpoints. In the anyon 
picture, the fusion of these endpoints corresponds to anyon--anti-anyon annihilation.
As we explain below, this fusion is also the 
X spider of ZX calculus.

In addition to the Frobenius algebra structure, $L^2(G)$ carries a second product 
corresponding to the stacking of open string worldsheets:
\begin{align}
    \mathtikz{\nablaA{0}{0}}
\end{align}
This is pointwise multiplication  on $L^{2}(G)$, which endows it with the structure of a Hopf algebra. 
According to the Peter-Weyl theorem,  this stacking can also be viewed as the entangled fusion of anyons  anti-anyon pairs: 
\begin{align}
    \mathtikz{\nablaA{0}{0}} = \bigoplus_{a,b,c} \vcenter{\hbox{ \includegraphics[scale=.06]{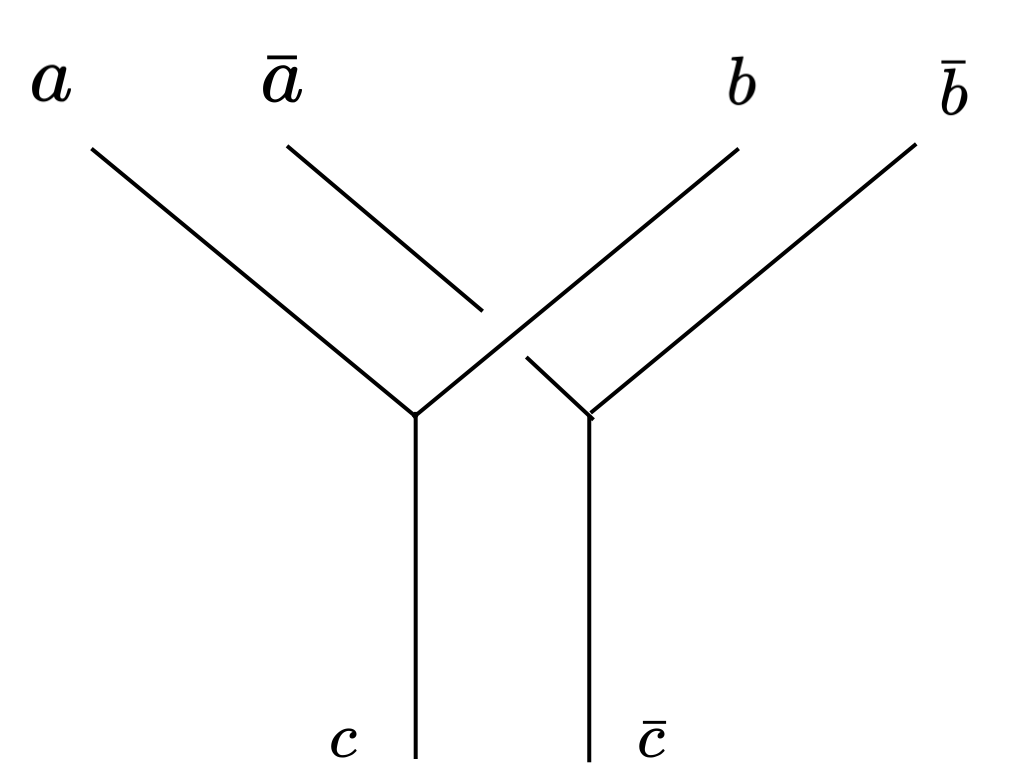}}} \times (\text{normalization})
\end{align}
Here, each term in the diret sum on the RHS is the projection of $V_{a}\otimes V_{\bar a} \otimes  V_{b}\otimes V_{\bar{b}}$ onto $V_{c}\otimes V_{\bar c}$: This can be computed explicitly using products of 3j symbols. 
Note that from the point of view of anyons this diagram involves a swap.

We will see that it maps on to the Z spider.  It was   introduced by \cite{Donnellynote} in the context of \cite{Donnelly:2020teo,Jiang:2020cqo}, where a large N, q deformed version of 2DYM was studied.  The q deformation makes this multiplication noncommutative.  From the anyon point of view this introduces braiding.

    \paragraph{Relation between the closed and open theory}
   In the continuum theory, it is important to relate the open and closed sector in a self consistent way. These two Frobenius algebras are related by algebra homomorphisms given by the zipper and co-zippers: 
\begin{align}
\mathtikz{\zipper{0}{0}} \mathtikz{\cozipper{0}{0}}
\end{align}
They satisfy a set of open-closed relations given by the Moore-Segal axioms.   This interplay between the closed and open sector is a new element relative to the conventional formulation of ZX calculus.  

A particularly important relation is  ``shrinkability":
\begin{align}
    \mathtikz{ \FrobetaA{0}{0} \cozipper{0}{0}} =\mathtikz{\etaC{0}{0}}
\end{align}
It was shown in \cite{Donnelly:2018ppr} that this condition implies that all holes created by the endpoints of an interval can be closed.  In particular, this implies that the Frobenius co-product is an isometry:
\begin{align}
    \mathtikz{\deltaA{0}{1cm} \muA{0}{0}} = \mathtikz{\idA{0}{0}}
\end{align}
This is a constraint on the boundary condition $e$ that is imposed on the endpoints of the interval.     \cite{Donnelly:2018ppr} showed that such a ``shrinkable" boundary condition exists in 2DYM and is given by the local condition:
\begin{align}
A_{t}|_{\text{bdry}}=0
\end{align}
where $t$ is the component of the gauge field along in the time direction.   This is the boundary condition which gave rise to $L^{2}(G)$ as the Hilbert space on the interval.   A similar shrinkable boundary condition can be defined in gauge theory in four dimensions \cite{Ball:2024hqe, Ball:2024gti}.

\section{ZX calculus for two dimensional gauge theory}
\subsection{Mapping to ZX}
\paragraph{$\mathbb{C}[G]$ and $L^{2}(G)$ for infinite $G$}
The ZX calculus was originally defined as a diagrammatic language for a finite group algebra.  In the context of two dimensional gauge theory, we want to generalize ZX to describe the group algebra of the gauge group, which is infinite. 
 
 When $G$ is finite, the group algebra $\mathbb{C}[G]$ is just the dual space of linear functionals on $L^{2}(G)$,  spanned by the group basis elements $\bra{U}$ for $U\in G$.  These act on a function $\ket{f} \in L^{2}(G)$ by evaluation:
\begin{align}
\braket{U|f}=f(U)
\end{align}
and have a multiplication defined by the group law:
\begin{align}
    \bra{U} \otimes  \bra{U'} \to \bra{U U'}.
\end{align}
For finite $G$, we can define the group basis $\ket{V}$ in $L^{2}(G)$ as delta functions. 
\begin{align}
    \braket{U|V} = \delta (U,V)
\end{align}
such that the group multiplication in $\mathbb{C}[G]$ corresponds to convolution product in $L^{2}(G)$.  

However, when $G$ is infinite, this isomorphism between $\mathbb{C}[G]$ and $L^{2}(G)$ no longer exists, because a delta function is not square integrable\footnote{We cannot even define $\bra{U}$ by evaluation, because two elements of $L^{2}(G)$  that differ on a set of measure zero are viewed as the same function.  But evaluation on these different functions yields differenct answers, so the evaluation map is strictly not well defined.   }.  Nevertheless, there is still an algebra on $L^{2}(G)$ defined by the convolution product, so we can take this to be what we mean by the group algebra. 

For the purpose of comparing with conventional definitions in ZX calculus, we will abuse notation  and continue to make use of the group basis $\ket{U}$ for infinite $G$ and expand functions in $L^{2}(G)$ as 
\begin{align}
    \ket{f} = \int dU f(U) \ket{U}.
\end{align}
In the infinite group context, we make sense of ``position basis" $\ket{U}$ by expanding it in the representation basis.  Then the fact that $\ket{U}$ is morally a delta function is reflected in the completeness of $a_{ij}(U)$:
\begin{align}
    \braket{U|V} \equiv \sum_{a, i,j } d_{a} a_{ij}(U) a^{*}_{ij}(V) =\delta(U,V)
\end{align}
For $G=U(1)$, this is just the usual Fourier transform of the delta function.  Similarly, we continue to use the notation  $\braket{U|f}\equiv f(U)$.
It will be convenient to define a normalized basis $\ket{a,i,j}$ of matrix elements, given by:
\begin{align}
    \braket{U|a,i,j} =\sqrt{d_{a}} a_{ij}(U) 
\end{align}
\paragraph{ZX for gauge theory}
ZX calculus describes operations on $\mathbb{C}[G]$  generated by the Z and X spiders, which we define as follows.  
The X spider  is the group product in $\mathbb{C}[G]$, which maps to the convolution product for $L^{2}(G)$.  It defines the product for a Frobenius algebra and is depicted by
\begin{align}
\input{x-spider-mult.tikz} \quad = \quad \input{x-spider-mult-ribbon.tikz} \quad : \ket{g} \otimes \ket{h} \to \ket{gh}.
\end{align}
The unit element is a delta function on the group, depicted as \begin{align}
    \begin{tikzpicture}
	\begin{pgfonlayer}{nodelayer}
		\node [style=rn] (0) at (0, 0.375) {};
		\node [style=none] (3) at (0, -0.625) {};
	\end{pgfonlayer}
	\begin{pgfonlayer}{edgelayer}
		\draw (0) to (3.center);
	\end{pgfonlayer}
\end{tikzpicture}
 \quad = \quad \input{x-unit-ribbon.tikz} \quad = \quad \ket{1},\qquad 
    \braket{g|1}= \delta(g,1) 
\end{align}
To represent a general basis state $\ket{g}$ where $g\in G$ is a group element, we simply label it in the diagram:
\begin{align}
\begin{tikzpicture}
	\begin{pgfonlayer}{nodelayer}
		\node [style={rn_phase}] (0) at (0, 0.375) {$g$};
		\node [style=none] (1) at (0, -0.625) {};
	\end{pgfonlayer}
	\begin{pgfonlayer}{edgelayer}
		\draw (0) to (1.center);
	\end{pgfonlayer}
\end{tikzpicture}
 \quad = \quad \input{x-state-ribbon.tikz} \quad = \quad \ket{g}
\end{align}
The Z spider is point wise multiplication on $L^{2}(G)$, which defines a Hopf algebra.
\begin{align}
\input{z-spider-mult.tikz} \quad = \quad \input{z-spider-mult-ribbon.tikz} \quad : \ket{g} \otimes \ket{h} \to  \delta_{g,h} \ket{g},
\end{align}
As alluded to earlier, this corresponds to the stacking of string worldsheets.
The Hopf algebra unit is the constant function. In the group algebra , this is an average over the group
\begin{align}
\begin{tikzpicture}
	\begin{pgfonlayer}{nodelayer}
		\node [style=gn] (0) at (0, 0.375) {};
		\node [style=none] (1) at (0, -0.625) {};
	\end{pgfonlayer}
	\begin{pgfonlayer}{edgelayer}
		\draw (0) to (1.center);
	\end{pgfonlayer}
\end{tikzpicture}
 \quad = \quad \input{z-unit-ribbon.tikz} \quad = \quad \int \ket{g} dg
\end{align}
Labeling the state with an arbitrary function $f \in L^{2}(G)$, we can represent the state $\ket{f}$:
\begin{align}
\begin{tikzpicture}
	\begin{pgfonlayer}{nodelayer}
		\node [style={gn_phase}] (0) at (0, 0.375) {$f$};
		\node [style=none] (1) at (0, -0.625) {};
	\end{pgfonlayer}
	\begin{pgfonlayer}{edgelayer}
		\draw (0) to (1.center);
	\end{pgfonlayer}
\end{tikzpicture}
 \quad = \quad \input{z-state-ribbon.tikz} \quad = \quad \int f(g) \ket{g} dg
\end{align}

More generally, we can define labelled Z and X spiders with arbitrary number of legs.
For any $n, m \in \mathbb{N}$, the Z spider with $n$ inputs and $m$ outputs is labelled by a function $f \in L^2(G)$ and is given by:
\begin{equation}
        \input{z-spider.tikz} \quad = \quad \input{z-spider-ribbon.tikz} 
        \quad \longmapsto \quad \int f(g) \ket{g}^{\otimes m} \bra{g}^{\otimes n} dg
\end{equation}

We will omit the label when it is the constant function $f(g)=1$.
Similarly, for any $n, m \in \mathbb{N}$, the X spider with $n$ inputs and $m$ outputs is labelled by a group element $g \in G$ and takes the form:
\begin{equation}
    \input{x-spider.tikz} \quad = \quad \input{x-spider-ribbon.tikz} 
    \quad \longmapsto \quad \int \delta\!\left(g h_1 \dots h_m (k_1 \dots k_n)^{-1} \right) \ket{h_1, \dots, h_m} \bra{k_1, \dots, k_n} d\vec{h}\, d\vec{k}
\end{equation}
We will omit the label when it is the identity element $g = 1$.
We are essentially taking ZX, thickening each line into  ribbon, and distinguishing the Z and X spider with 2D topoloogy.   Many of the  usual ZX rules apply, but some of the re-write rules are now manifest from 2D topology.   We give a list of these re-writes below.   The soundness of  these diagrams follows from interpreting the  ribbon graphs as entangled  anyon worldlines, which follows the standard rules of modular tensor categories.

\subsection{Relating ribbon to anyon diagrams}
To derive the ZX re-write rules diagrammatically, it will be useful make use of both the worldsheet and anyon interpretation of the ribbons.  For this purpose,  we provide a precise dictionary between the ribbon graphs and anyon diagrams below.  The latter is best described in the representation basis, with $\ket{a,i,j} =\ket{a,i}\otimes \ket{\bar a,j} $ viewed as entangled states of  anyon and anti-anyons.

The unit and convolution product in the Frobenius algebra are 
\begin{align}
\input{x-unit-ribbon.tikz}&=\sum_{a} \sqrt{d_{a}}
\vcenter{\hbox{
\includegraphics[scale=.06]{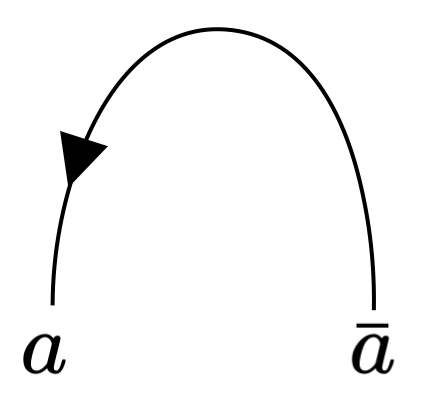}}}=\sum_{a,i}e^{-A C_{2}(a)} \sqrt{d_{a}} \ket{a, i,i}\nn
\input{x-spider-mult-ribbon.tikz}&=\sum_{a} \frac{1}{\sqrt{d_{a}}} \vcenter{\hbox{
\includegraphics[scale=.15]{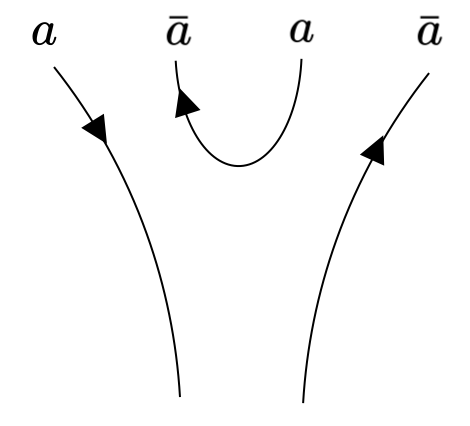}}}=\sum_{a}  \sum_{R,i,,j,k} \frac{e^{-A C_{2}(R)}}{\sqrt{d_{a}}} \ket{a ,i ,k} \bra{a,i,j}\bra{a,j,k},
\end{align}
while the co-unit and the co-product are 
obtained by flipping the diagrams upside down and turning kets into bras.  The Boltzmann factor $e^{-AC_{2}(a)}$ involve the area $A$ of the ribbons and the quadratic Casimir $C_{2}(a)$, which provides a convergence factor for the infinite sum over $a$.   In this language, the isometry condition is equivalent to the fact that anyon loops evaluate to $d_{a}$.

The unit and the point wise multiplication corresponds to the trivial anyon and fusion
\begin{align}
    \input{z-unit-ribbon.tikz} &=  \vcenter{\hbox{ \includegraphics[scale=.1]{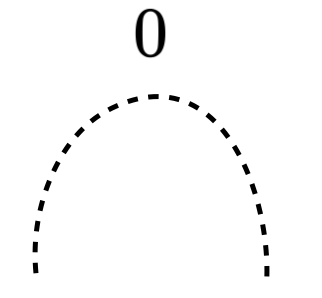}}}   = 1 \nn
    \input{z-spider-mult-ribbon.tikz} &= \bigoplus_{a,b,c} \vcenter{\hbox{ \includegraphics[scale=.06]{figures/entangled_fusion.png}}} \times (\text{normalization})\nn
    &=\sum_{\substack{a,b,c \\ i,j,k \\ l,m,n}}  \begin{pmatrix} a&&b&&c\\
    i&&j&&k
    \end{pmatrix} \begin{pmatrix} \bar a&&\bar b&&\bar c\\
    l&&m&&n  \end{pmatrix} \ket{c ,k,m} \bra{a,i,j}\bra{a,i,l}\bra{b,j,m},
\end{align}
Here $\begin{pmatrix} a&&b&&c\\
    i&&j&&k
    \end{pmatrix} \begin{pmatrix} \bar a&&\bar b&&\bar c\\
    l&&m&&n  \end{pmatrix} = \int dU \,\,  a_{il}(U) b_{jm}(U) c^*_{kl}(U)$ are products of Clebsch Gordon coefficients (3j symbols).
\subsection{ZX re-write rules} 
Here we present the ZX re-write rules and explain how the ribbon diagrams make them topologically manifest.  

\paragraph{Identity}
The one-input one-output Z and X spiders are equal to the identity.
For the Z spider, this corresponds to removing the trivial anyon in the ribbon diagram; while for the X spider, this equality is trivial.
\begin{equation}
    \input{z-x-id.tikz} \quad \longleftrightarrow \quad \input{z-x-id-ribbon.tikz}
\end{equation}

\paragraph{Units}
The simplest re-write is the statement that $\begin{tikzpicture}
	\begin{pgfonlayer}{nodelayer}
		\node [style=gn] (0) at (0, 0.375) {};
		\node [style=none] (1) at (0, -0.625) {};
	\end{pgfonlayer}
	\begin{pgfonlayer}{edgelayer}
		\draw (0) to (1.center);
	\end{pgfonlayer}
\end{tikzpicture}
 $ and  $\begin{tikzpicture}
	\begin{pgfonlayer}{nodelayer}
		\node [style=rn] (0) at (0, 0.375) {};
		\node [style=none] (3) at (0, -0.625) {};
	\end{pgfonlayer}
	\begin{pgfonlayer}{edgelayer}
		\draw (0) to (3.center);
	\end{pgfonlayer}
\end{tikzpicture}
$ are unit elements.
\begin{equation}
\input{z-mult-unit.tikz}
\end{equation}
In the ribbon language:
\begin{align}
\input{z-mult-unit-ribbon.tikz}  = \begin{tikzpicture}
	\begin{pgfonlayer}{nodelayer}
		\node [style=none] (0) at (-0.5, -1) {};
		\node [style=none] (1) at (0.5, -1) {};
		\node [style=none] (2) at (-0.5, 1) {};
		\node [style=none] (3) at (0.5, 1) {};
	\end{pgfonlayer}
	\begin{pgfonlayer}{edgelayer}
		\draw (1.center)
			 to (0.center)
			 to (2.center)
			 to (3.center)
			 to cycle;
	\end{pgfonlayer}
\end{tikzpicture}
 \longleftrightarrow
\sum_{a,b,c}
\vcenter{\hbox{\includegraphics[scale=.2]{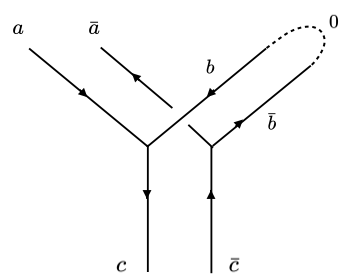}}}=\sum_{a}\vcenter{\hbox{\includegraphics[scale=.2]{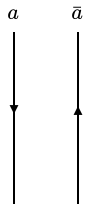}}}
\end{align}
In the worldsheet picture, the z-unit deletes a sheet, while in the anyon picture it is the vacuum anyon and force the $b$ anyon to be trivial.
For the $X$ unit, we have
\begin{equation}
\input{x-mult-unit.tikz}
\end{equation}
In the ribbon language:
\begin{align}
\input{x-mult-unit-ribbon.tikz}  = \begin{tikzpicture}
	\begin{pgfonlayer}{nodelayer}
		\node [style=none] (0) at (-0.5, -1) {};
		\node [style=none] (1) at (0.5, -1) {};
		\node [style=none] (2) at (-0.5, 1) {};
		\node [style=none] (3) at (0.5, 1) {};
	\end{pgfonlayer}
	\begin{pgfonlayer}{edgelayer}
		\draw (1.center)
			 to (0.center)
			 to (2.center)
			 to (3.center)
			 to cycle;
	\end{pgfonlayer}
\end{tikzpicture}

\end{align}

\paragraph{Convolution}
The following rule captures diagrammatically that the X-spider is a convolution produce on $L^2(G)$: 
\begin{equation}
    \input{convolution.tikz} \quad \longleftrightarrow \quad \input{convolution-ribbon.tikz}
\end{equation}

\paragraph{Fusion}
An essential part of the ZX-calculus involves fusion of the X and Z spiders.
These correspond to the standard Frobenius relations of a Frobenius algebra, generalized to include labels.
The fusion for the Z spider is shown below:
\begin{equation}
    \input{z-fusion.tikz} \quad \longleftrightarrow \quad \input{z-fusion-ribbon.tikz},
\end{equation}
Similarly, we have the fusion rule for the X spider.
\begin{equation}
    \input{x-fusion.tikz} \quad \longleftrightarrow \quad \input{x-fusion-ribbon.tikz}
\end{equation}
\begin{equation}
    \input{x-fusion-2.tikz} \quad \longleftrightarrow \quad \input{x-fusion-ribbon-2.tikz}
\end{equation}
Note that the non-commutativity of the X spider means that the labels can only combine when they are connected along the boundary of a ribbon.
Meanwhile, the commutativity of the Z ribbon is expressed as:
\begin{equation}
    \input{z-symmetry.tikz}
\end{equation}
\begin{equation}
    \input{z-symmetry-ribbon.tikz}
\end{equation}

\paragraph{Bi-algebra rules}
    The bi-algebra relations are best understood in the worldsheet picture. The first one says that says that zero-strings splitting produces zero-strings. 
\begin{align}
    \input{z-copy.tikz} \quad  \longleftrightarrow \quad \input{z-copy-ribbon.tikz}
\end{align}
The next says that we can peel apart the co-unit
\begin{equation}
\input{x-copy.tikz}  \quad  \longleftrightarrow  \quad \input{x-copy-ribbon.tikz}
\end{equation}
\begin{equation}
    \input{x-push.tikz} \quad  \longleftrightarrow \quad \input{x-push-ribbon.tikz}
\end{equation}
Similarly, the bi-algebra says we can peel apart the co-product:
\begin{equation} 
    \input{bialgebra.tikz} \quad \longleftrightarrow \quad \input{bialgebra-ribbon.tikz}
\end{equation}

\paragraph{Hopf algebra rules}

A Hopf algebra has an antipode.  On $L^{2}(G)$
\begin{align}
    S: f(g) \to  f(g^{-1})
\end{align}
In ZX-calculus, this is represented by the composition of a Z cup and an X cap (and vice versa):
\begin{equation}
    \input{antipode.tikz}
\end{equation}
We can use this to derive that the antipode corresponds to a 180 degree twist of the ribbon:
\begin{equation}
    \input{antipode-ribbon.tikz}
\end{equation}
This flips the orientation of strings.
If we input the state $\ket{g}$ into the antipode, we can see that the twist inverts the group element:
\begin{equation}
    \input{antipode-state-inverse.tikz}
\end{equation}
The defining identity of the anti-pode is the Hopf rule:
\begin{equation}
    \input{antipode-hopf.tikz}
    \quad \longleftrightarrow \quad
    \input{antipode-hopf-ribbon.tikz}
\end{equation}
In 2DYM it corresponds to the identity $U U^{-1} = U^{-1}U = 1$.
This is not obvious from the string worldsheet point of view, but it becomes manifest in the interpretation in terms of entangled anyon worldlines:
\begin{align}
\sum_{a} \frac{1}{d_{a}}\vcenter{\hbox{\includegraphics[scale=.1]{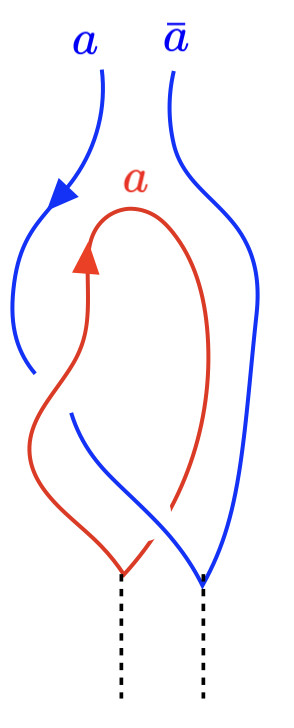}}} = \sum_{a} \vcenter{\hbox{\includegraphics[scale=.1]{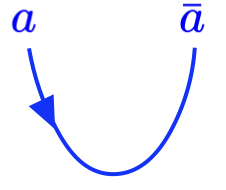}}} 
\end{align}

\section{Conclusion}
In this work we have described a ribbon ZX-calculus for 2DYM with compact gauge group.  We have found it useful to interpret the  ribbons as entangled anyons, which obey the fusion rules of a compact group, but have trivial braiding.   A natural generalization would be to consider  q-deformed Yang-Mills theory associated to a quantum group, where the ribbons would obey nontrivial braiding rules.  Indeed, the study of q2DYM \cite{Donnellynote} was an essential motivation behind this work.  Braided ZX calculus has also been studied by \cite{Majid_2022}. 

Another important generalization in the context of high energy theory involves the taking the large N limit.  For $U(N)$, this leads to the Gross-Taylor string theory \cite{Gross:1993hu}, whose ribbon diagrammatics was studied in \cite{Donnelly:2018ppr,Donnelly:2016jet}. 
An even richer physical system arises when we introduce a combination of  q deformation and large N : this leads to a topological string theory, with the $q$ deformation parameter  determined by the string coupling.  This system is ``gravitational"  in the sense that it exhibits backreaction of D branes to geometry, and has a close relationship to physical string theories which are full fledged quantum theories of gravity.  

A further motivation behind this work is to provide the first step in formulating a ZX calculus for low dimensional gravity.   The starting point for such a connection is two dimensional BF gauge theory, which satisfies the same diagrammatics as 2DYM, but with the area dependence replaced by a dependence on the boundary length of the two manifold.  As explained in  \cite{Blommaert:2018iqz},  this is equivalent to two dimensional JT gravity, provided that we choose the gauge group to be SL$(2,\mathbb{R})^+$, the positive semi-group of SL$(2,\mathbb{R})$ matrices with positive entries.  $q$ deforming this group (with q determined by the gravitational coupling) leads to a quantum semi-group structure underlying TQFT's that describe three dimensional quantum gravity \cite{Mertens:2022ujr} \footnote{  More precisely, pure 3d gravity is described by the representation category of SL$(2,\mathbb{R})_{q}^+ \otimes $ SL$(2,\mathbb{R})_{q}^+ $}.    We expect that generalizing our work to these more exotic group algebras will lead to a ZX calculus for spacetime.  Finally, we observe that the ribbon diagrammatics may also be useful in the context of short range entangled, many body systems described by an SPT phase. In 1+1 D, this defines a ``computational phase" of matter that supports measurement based quantum computation (MBQC). In \cite{Wong:2023bhs} , it was shown that the diagrammatics of 
 b- twisted group algebra\footnote{ In  \cite{Wong:2023bhs}, these b twisted group algebras appeared in the context of equivariant TQFT} $\mathbb{C}_{b}(G)$ provides a useful description for the global aspects of MBQC. 
\section*{Acknowledgements}
GW is supported by STFC grant
ST/X000761/1 and the Oxford Mathematical Institute.
RS is supported by the Clarendon Fund Scholarship.
\appendix

\section*{Appendix}
\section{Hopf algebras and 2D Yang-Mills}

2D Yang-Mills can be cast as an open-closed TQFT whose Hilbert space is a space of functions on a group.
This space has the structure of a Hopf algebra.


\subsection{Bialgebra}

On our algebra $H$ of functions on a group $G$ we define operations:
\begin{align}
\text{product} \qquad \nabla &: H \otimes H \to H \\
\text{unit} \qquad \eta &: \mathbb{C} \to H \\
\text{coproduct} \qquad \Delta &: H \to H \otimes H \\
\text{counit} \qquad \epsilon &: H \to \mathbb{C}
\end{align}
These define a bialgebra structure.
It is also useful to introduce the element
\begin{equation}
\tau : H \otimes H \to H \otimes H
\end{equation}
defined so that $\tau(f \otimes g) = g \otimes f$.

The compatibility condition of the unit with the product is
\begin{equation}
\nabla \circ (1 \otimes \eta) = \nabla \circ (\eta \otimes 1) = 1
\end{equation}
and of the counit with the coproduct is
\begin{equation}
(1 \otimes \epsilon) \circ \Delta = (\epsilon \otimes 1) \circ \Delta = 1.  
\end{equation}

There are compatibility conditions between the product and coproduct structures:
\begin{align}
\Delta \circ \nabla &= (\nabla \otimes \nabla) \circ (1 \otimes \tau \otimes 1) \circ (\Delta \otimes \Delta) \\
\epsilon \circ \nabla &= \epsilon \otimes \epsilon \\
\Delta \circ \eta &= \eta \otimes \eta \\
\epsilon \circ \eta &= 1.
\end{align}

Let $f,g$ denote elements of $H$, i.e. functions defined on $g$. Let $u,v$ denote elements of $G$ and $id$ denote the identity element.
Then the definitions are
\begin{align}
\nabla(f,g)(u) &= f(u) g(u) \\
\eta(u) &= 1 \\
\Delta(f)(u,v) &= f(uv) \\
\epsilon(f) &= f(id).
\end{align}
It's a straightforward exercise to see that these satisfy the axioms of a bialgebra.
It is commutative ($\nabla \circ \tau = \nabla$) but not cocommutative ($\Delta \circ \tau \neq \Delta$).
 
In 2D Yang-Mills the coproduct $\Delta$ is used in the splitting of the Hilbert space of an interval into two intervals, it corresponds to the cutting of strings.
Similarly, the counit is used in the definition of the E-brane state.

The product $\nabla$ is just pointwise multiplication of functions on the group.
We use this implicitly in the string description when we construct states of multiple strings; the multi-string state $\ket{IJ}$ is really just a bunch of single string states $\ket{ij}$ combined via the product.
The corresponding unit for this multiplication is the function $1(x) = 1$. In terms of strings, this is the vacuum state containing no strings.
This structure is not changed under $q$-deformation.

\subsection{Hopf algebra}

The Hopf algebra has an additional operation, the antipode $S: H \to H$ which is the pullback of the inverse on the group:
\begin{equation}
S(f)(u) = f(u^{-1}).
\end{equation}
It satisfies 
\begin{equation} \label{Sdef}
\nabla \circ (S \otimes 1) \circ \Delta = \eta \circ \epsilon.
\end{equation}
This defining relation implies some further properties, c.f. exercise 1.3.3 of \cite{Majid:1990vz}:
\begin{align}
S(ab) &= S(b) S(a), \label{Sprop1} \\
(S \otimes S) \circ \Delta &= \tau \circ \Delta \circ S, \label{Sprop2} \\
\epsilon \circ S &= \epsilon. \label{Sprop3} 
\end{align}

In terms of Yang-Mills, this is flipping the orientation: it turns a right-oriented string into a left-oriented string.

\bibliographystyle{utphys}
\bibliography{hopf}

\end{document}